\documentclass[useAMS]{mn2e}

\usepackage{latexsym,graphicx}

\usepackage{color}

\newcommand\kms{{\rm\,km\,s^{-1}}}
\newcommand\msun{\rm\,M_\odot}
\newcommand\lsun{\rm\,L_\odot}
\newcommand\rsun{\rm\,R_\odot}

\def\apgt{\ {\raise-.5ex\hbox{$\buildrel>\over\sim$}}\ }
\def\aplt{\ {\raise-.5ex\hbox{$\buildrel<\over\sim$}}\ }

\title[TYC\,3159-6-1: a runaway blue supergiant]{TYC\,3159-6-1: a runaway blue supergiant}
\author[V.V.Gvaramadze et al.]
        {V. V.~Gvaramadze,$^{1,2}$\thanks{E-mail: vgvaram@mx.iki.rssi.ru}
        A. S.~Miroshnichenko,$^{3}$ N.~Castro,$^{4}$
        N.~Langer$^{4}$
        \newauthor
        and S. V.~Zharikov$^{5}$\\
        $^{1}$Sternberg Astronomical Institute, Lomonosov Moscow State University, Universitetskij Pr. 13, Moscow 119992, Russia\\
        $^{2}$Isaac Newton Institute of Chile, Moscow Branch, Universitetskij Pr. 13, Moscow 119992,
        Russia \\
        $^{3}$Department of Physics and Astronomy, University of North Carolina at Greensboro, Greensboro, NC 27402-6170, USA \\
        $^{4}$Argelander-Institut f\"ur Astronomie der Universit\"at Bonn, Auf dem H\"ugel 71, 53121, Bonn, Germany \\
        $^{5}$Instituto de Astronom\'{a}, Universidad Nacional Aut\'{o}noma de M\'{e}xico, Ensenada, Mexico\\
               }
\begin{document}

\date{Accepted 2013 October 24.  Received 2013 October 24; in original form 2013 July 26}

\maketitle

\label{firstpage}

\begin{abstract}
We report the results of optical spectroscopy of a candidate
evolved massive star in the Cygnus\,X region, TYC\,3159-6-1,
revealed via detection of its curious circumstellar nebula in
archival data of the {\it Spitzer Space Telecope}. We classify
TYC\,3159-6-1 as an O9.5$-$O9.7\,Ib star and derive its
fundamental parameters by using the stellar atmosphere code {\sc
fastwind}. The He and CNO abundances in the photosphere of
TYC\,3159-6-1 are consistent with the solar abundances, suggesting
that the star only recently evolved off the main sequence. Proper
motion and radial velocity measurements for TYC\,3159-6-1 show
that it is a runaway star. We propose that Dolidze\,7 is its
parent cluster. We discuss the origin of the nebula around
TYC\,3159-6-1 and suggest that it might be produced in several
successive episodes of enhanced mass-loss rate (outbursts) caused
by rotation of the star near the critical, $\Omega$-limit.
\end{abstract}


\section{Introduction}
\label{sec:int}

A significant fraction of massive stars leave the confines of
their parent clusters because of few-body dynamical encounters
with other cluster's members or binary-supernova explosions, and
spread out into the Galactic disk and halo to form the population
of field stars. Some of the field stars possess high space
velocities (the so-called runaway stars; Blaauw 1961) and can
reach large distances from their birthplaces. The runaway stars
escape from the large-scale wind bubbles, created around their
parent clusters by the cumulative effect of stellar winds and
supernovae, and their individual wind bubbles transform into bow
shocks (Weaver et al. 1977). Depending on the physical properties
(temperature, number density) of the ambient interstellar medium
(ISM), the bow shocks might be generated and observable during
most of the lifetime of runaway stars (Huthoff \& Kaper 2002).
This makes them the most wide-spread sort of parsec-scale nebulae
associated with the field OB stars (e.g. van Buren, Noriega-Crespo
\& Dgani 1995; Peri et al. 2012).

The infrared (IR) surveys carried out by the {\it Spitzer Space
Telescope} (Werner et al. 2004) and the {\it Wide-field Infrared
Survey Explorer} ({\it WISE}; Wright et al. 2010) greatly
increased the number of known bow shocks. Also very important is
that they revealed numerous compact nebulae of various
morphologies, which are reminiscent of circumstellar nebulae
observed around evolved massive stars (Gvaramadze, Kniazev \&
Fabrika 2010; Wachter et al. 2010; Mizuno et al. 2010). Follow-up
spectroscopy of central stars of these nebulae resulted in the
discovery of a large number of massive stars at the blue
supergiant, luminous blue variable (LBV) and Wolf-Rayet stages
(e.g. Gvaramadze et al. 2012a and earlier papers; Wachter et al.
2010). The short duration of these transient phases in the life of
massive stars (a factor of 10 to 100 shorter than the
main-sequence phase) implies that their circumstellar nebulae are
rare objects. On the other hand, the rich morphological diversity
of these nebulae (from circular to bipolar and triple-ring shape)
points to the existence of several mechanisms responsible for
their production and shaping. The detection of new examples of
circumstellar nebulae allow us to not only reveal evolved massive
stars, but it might also be crucial for better understanding the
origin of these nebulae and the post-main-sequence evolution of
their central stars.

In this paper, we report the discovery of a blue supergiant star
in the Cygnus\,X region through detection of its associated nebula
in the archival data of the {\it Spitzer Space Telescope} and
follow-up optical spectroscopy. The star, TYC\,3159-6-1, and the
nebula are presented in Section\,\ref{sec:neb}. The spectral
classification of TYC\,3159-6-1 and the results of modelling of
its spectrum with the stellar atmosphere code {\sc fastwind} are
given in Section\,\ref{sec:spec}. In Section\,\ref{sec:dis}, we
show that TYC\,3159-6-1 is a runaway star, whose likely birthplace
is the star cluster Dolidze\,7. The possible origin of the nebula
around TYC\,3159-6-1 is discussed in the same section.

\section{TYC\,3159-6-1 and its associated mid-infrared nebula}
\label{sec:neb}

The nebula around TYC\,3159-6-1 was discovered during our search
for evolved massive stars around the Cyg\,OB2 association via
detection of their circumstellar nebulae (e.g. Gvaramadze et al.
2009, 2010). The search was carried out by using archival data
originating from the Cygnus-X {\it Spitzer} Legacy Survey (Hora et
al. 2008)\footnote{http://www.cfa.harvard.edu/cygnusX}. This
survey covers 24 square degrees in Cygnus\,X, one of the most
massive star-forming complexes in the Milky Way (e.g., Reipurth \&
Schneider 2008), and provides images at 24 and 70\,$\mu$m obtained
with the Multiband Imaging Photometer for {\it Spitzer} (MIPS;
Rieke et al. 2004) and at 3.6, 4.5, 5.8, and 8.0\,$\mu$m obtained
with the Infrared Array Camera (IRAC; Fazio et al. 2004). The
resolution of the MIPS 24 and 70\,$\mu$m images is $\approx 6$ and
18 arcsec, respectively, while that of the IRAC images is $\approx
1$ arcsec.

\begin{figure}
\includegraphics[width=8cm]{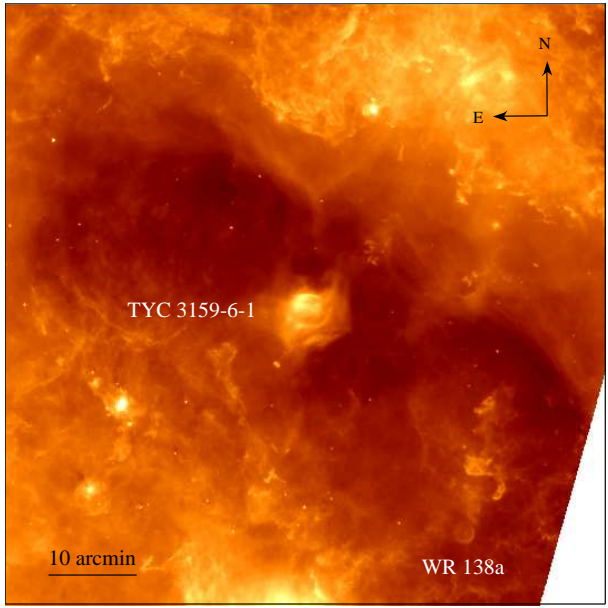}
\centering \caption{MIPS 24\,$\mu$m image of the field containing
two IR nebulae associated with TYC\,3159-6-1 (the subject of this
paper) and the WN8-9h star WR\,138a.} \label{fig:WR138a}
\end{figure}

Visual inspection of the 24\,$\mu$m survey data have led to the
discovery of several compact nebulae. One of them turns out to be
associated with the curious evolved massive star MWC\,349A
(Gvaramadze \& Menten 2012). Two other nebulae, MN114 and
MN115\footnote{In the SIMBAD database these nebulae are named
[GKF2010] MN114 and [GKF2010] MN115.} (Gvaramadze et al. 2010),
are produced by the WN8-9h star WR\,138a (Gvaramadze et al. 2009)
and a candidate LBV star (Gvaramadze et al., in preparation),
respectively. Another nebula, which is the subject of this paper,
is associated with a moderately bright ($V\sim11$ mag) star,
located at only $\approx 0.5\degr$ to the northeast of WR\,138a
and its ring nebula (see Fig.\ref{fig:WR138a} for the MIPS
24\,$\mu$m image of the field containing both nebulae). Subsequent
examination of the SIMBAD data
base\footnote{http://simbad.u-strasbg.fr/simbad/} revealed that
this IR nebula was independently discovered by Takita et al.
(2009) by using the 9 and 18\,$\mu$m images of the Cygnus X region
taken as part of the {\it AKARI} mid-IR All-Sky Survey. These
authors also identified a point-like source in the centre of the
nebula with an optically visible star, listed in the Tycho catalog
(Egret et al. 1992) as TYC\,3159-6-1.

At 9\,$\mu$m the {\it AKARI} image shows an IR counterpart to
TYC\,3159-6-1 and a weak diffuse emission to the northeast of the
star. At 18\,$\mu$m TYC\,3159-6-1 is almost hidden in the bright
emission of the nebula, which appears as a shell-like structure
centred on TYC\,3159-6-1 and two concentric, evenly spaced
arc-like structures to the southwest of the star.

To determine the nature of TYC\,3159-6-1, Takita et al. (2009)
carried out follow-up optical spectroscopy of this star on 2007
September 21 using the 1.5-m telescope at the Gunma Astronomical
Observatory. Although the obtained spectrum revealed the presence
of the H$\alpha$ emission line, with an equivalent width (EW) of
$-$1.5 \AA, the spectral classification of the star was not
carried out because of the low resolution ($R\sim 400-500$) of the
spectroscopic material. Takita et al. (2009) considered three
possibilities for the nature of TYC\,3159-6-1: a nearby low-mass
star, a highly reddened Be star, and an asymptotic giant branch
star with a detached shell. They concluded, however, that none of
them ``can account for the observation consistently". In
Section\,\ref{sec:clas}, we show that TYC\,3159-6-1 is an
O9.5$-$O9.7\,Ib star.

Using the observed 18\,$\mu$m flux of the nebula around
TYC\,3159-6-1, Takita et al. (2009) estimated the mass of the
nebula to be $\sim 0.04 \,  (d/1 {\rm kpc}) \, \msun$, where $d$
is the distance to the star.

\begin{figure*}
\includegraphics[width=15cm]{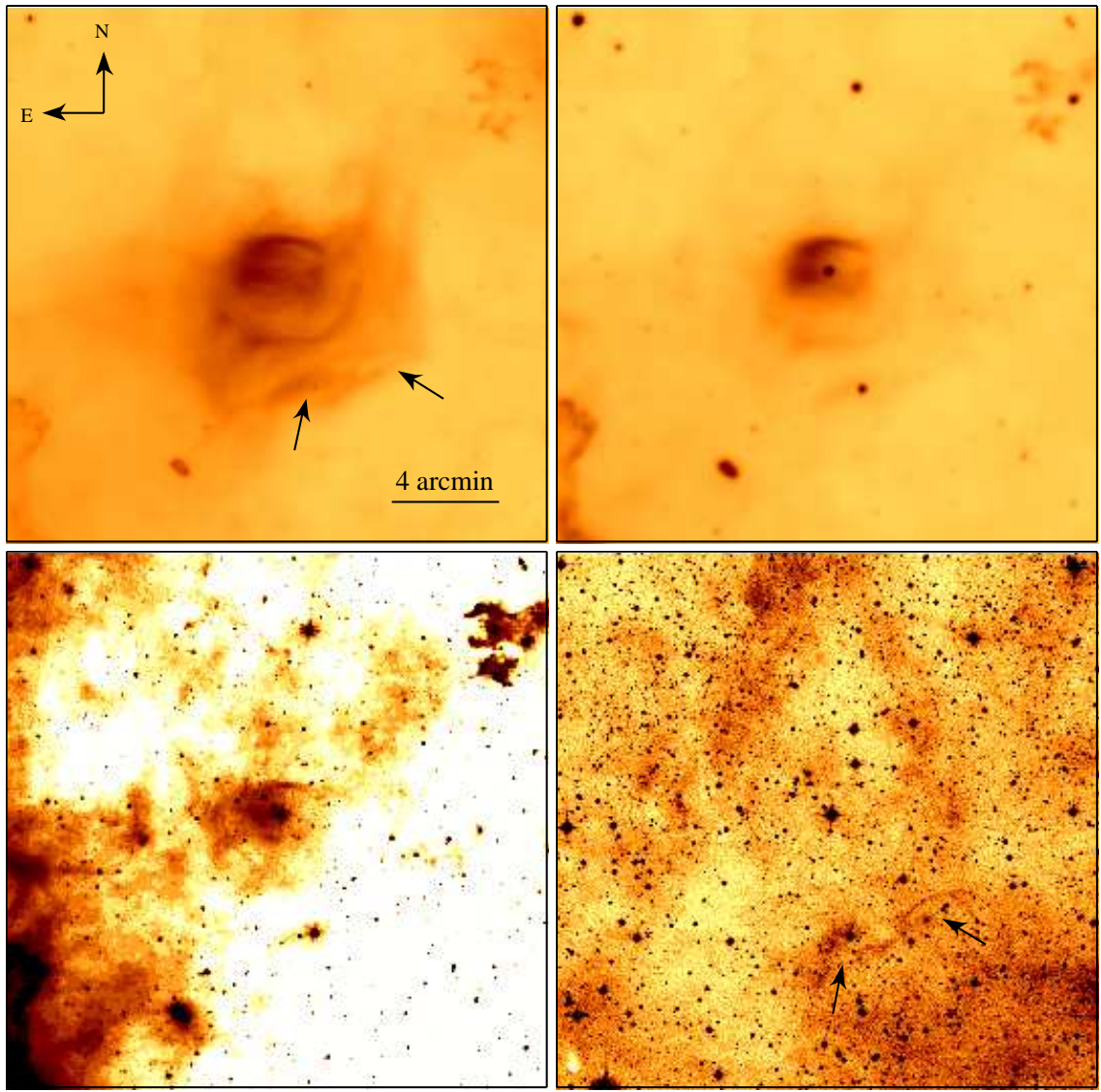}
\centering \caption{From left to right, and from top to bottom:
MIPS 24\,$\mu$m, {\it WISE} 12\,$\mu$m, IRAC 8\,$\mu$m, and DSS-II
red band images of the field containing TYC\,3159-6-1 and its
circumstellar nebula. Arrows in the MIPS and DSS-II images point
to filamentary structures visible both in the infrared and optical
bands. The orientation and the scale of the images are the same.
At a distance of 1.5 kpc, 1 arcmin corresponds to $\approx 0.43$
pc.} \label{fig:neb}
\end{figure*}
\begin{table}
  \caption{Details of TYC\,3159-6-1.}
  \label{tab:det}
  \begin{center}
 \begin{tabular}{lrr}
    \hline
  Spectral type & O9.5$-$O9.7\,Ib \\
  RA(J2000) & $20^{\rm h} 18^{\rm m} 40\fs37$  \\
  Dec(J2000) & $41\degr 32\arcmin 45\farcs0$ \\
  $l$ & $78\fdg8320$ \\
  $b$ & $3\fdg1505$  \\
  $B$ (mag) & 12.51$\pm$0.28 \\
  $V$ (mag) & 10.88$\pm$0.07 \\
  $I$ (mag) & 8.52$\pm$0.06 \\
  $J$ (mag) & 6.82$\pm$0.02 \\
  $H$ (mag) & 6.32$\pm$0.04 \\
  $K_{\rm s}$ (mag) & 5.95$\pm$0.02 \\
  $[3.4]$ (mag) & 5.78$\pm$0.05 \\
  $[4.6]$ (mag) & 5.47$\pm$0.03 \\
  $[12]$ (mag) & 5.33$\pm$0.02 \\
  $[22]$ (mag) & 3.41$\pm$0.12 \\
  \hline
 \end{tabular}
\end{center}
\end{table}

In Fig.\,\ref{fig:neb} we present 24, 12 and 8\,$\mu$m images of
the field containing TYC\,3159-6-1 and its circumstellar nebula,
obtained with the MIPS, {\it WISE} and IRAC, respectively, and the
Digitized Sky Survey II (DSS-II) red band (McLean et al. 2000)
image of the same field. The MIPS 24\,$\mu$m image shows a bright,
slightly asymmetric, incomplete shell of radius of $\sim 1.4$
arcmin, centred on a well-discerned point source -- TYC\,3159-6-1,
and an arc-like structure with a radius of $\approx 2.5$ arcmin to
the southwest of the star. One can also see a more diffuse
emission around the star (mostly concentrated to the southwest) of
radius $\sim 4$ arcmin and two filaments extending for $\sim 5$
arcmin to the northwest. One of these filaments apparently crosses
the bright shell and then curves to the northeast to form an
incomplete circle of radius $\sim 3$ arcmin to the north of
TYC\,3159-6-1. This incomplete circle is also discernible in the
{\it WISE} 12\,$\mu$m image and its east boundary is clearly
visible in the IRAC 8\,$\mu$m image. Moreover, the DSS-II image
shows that the circle is flanked from the east and west sides by
diffuse optical emission. The DSS-II image also shows distinct
optical counterparts to filamentary structures at the southwest
edge of the 24\,$\mu$m nebula (indicated by arrows in both
images).

The details of TYC\,3159-6-1 are summarized in
Table\,\ref{tab:det}. The coordinates and the $B$ and $V$
magnitudes are from the Tycho-2 Catalogue (H$\o$g et al. 2000).
The $J$, $H$ and $K_{\rm s}$ magnitudes are taken from the 2MASS
(Two Micron All Sky Survey) All-Sky Catalog of Point Sources
(Cutri et al. 2003). For the sake of completeness we also give the
$I$ magnitude from the Amateur Sky Survey (TASS; Droege et al.
2006) and the {\it WISE} 3.4, 4.6, 12 and 22 $\mu$m magnitudes
from the {\it WISE} Preliminary Release Source Catalog (Cutri et
al. 2012).

\section{TYC\,3159-6-1: a blue supergiant}
\label{sec:spec}

To determine the spectral type of TYC\,3159-6-1 and thereby to
estimate the distance to this star, we observed TYC\,3159-6-1
within the framework of our programme of spectroscopic follow-up
of candidate massive stars revealed via detection of their
circumstellar shells and bow shocks (e.g. Gvaramadze et al. 2011,
2012a, 2013; Stringfellow et al. 2012).

\subsection{Spectroscopic observations and data reduction}
\label{sec:obs}

We obtained two spectra of TYC\,3159-6-1.

The first one was taken on 2012 November 14 with the 2.1-m
telescope of the observatory of San Pedro Martir (Baja California,
Mexico) using the REOSC Espresso \'echelle spectrograph (Levine \&
Chakrabarty 1995). This instrument gives a resolution of 0.12
\AA\,pixel$^{-1}$ near H$\alpha$ (that corresponds to a spectral
resolving power of $\sim$20000) using a UCL camera and a CCD-E2V
chip of 2048$\times$2048 pixels with a 13.5 micron pixel size. The
spectra cover 28 orders and span the spectral range 3720--7200
\AA. One 20-minute exposure resulted in a range of signal-to-noise
ratios (SNRs) from $\sim$20 near H$\delta$ to $\sim$80 near
H$\alpha$.

The second spectrum was obtained on 2012 December 28 with the
2.7-m Harlan J. Smith telescope of the McDonald Observatory using
a coud\'e \'echelle spectrograph (Tull et al. 1995) that provides
a spectral resolving power of 60000 in the range 3600--10500 \AA.
One 15-minute exposure resulted in a range of SNRs from $\sim$15
near H$\delta$ to over 100 redward of H$\alpha$.

Both spectra were reduced using the {\sc iraf}\footnote{{\sc
iraf}: the Image Reduction and Analysis Facility is distributed by
the National Optical Astronomy Observatory (NOAO), which is
operated by the Association of Universities for Research in
Astronomy, Inc. (AURA) under cooperative agreement with the
National Science Foundation (NSF).} software and re-binned for
improving the SNR without affecting the spectral resolution. EWs,
FWHMs and heliocentric radial velocities (RVs) of main lines in
the spectrum of TYC\,3159-6-1 are summarized in
Table\,\ref{tab:lines}. The accuracies of EW, FWHM and RV
measurements are $\pm$(0.02$-$0.03) \AA, $\pm$0.1 \AA \, and
$\pm$$1 \, \kms$, respectively. Using the He\,{\sc i} lines, we
derived the heliocentric radial velocity of TYC\,3159-6-1 of
$v_{\rm r, hel}=-35.8\pm3.0 \, \kms$.

\begin{table}
\centering{ \caption{Equivalent widths (EWs), FWHMs and radial
heliocentric velocities (RVs) of main lines in the spectrum of
TYC\,3159-6-1 taken on 2012 November 14. The negative value of EW
of the H$\alpha$ line corresponds to its emission component. For
the sake of comparison, we also provide EWs (in brackets) of the
components of the H$\alpha$ line from the spectrum taken on 2012
December 28.} \label{tab:lines}
\begin{tabular}{lllc}
\hline 
$\lambda_{0}$(\AA) Ion  & EW($\lambda$) & FWHM($\lambda$) & RV \\
       & (\AA) & (\AA) & ($\kms$) \\
\hline 
4471\ He\,{\sc i}       & 0.73 & 2.60 & $-$38.7 \\
4713\ He\,{\sc i}       & 0.37 & 2.40 & $-$36.7 \\
4861\ H$\beta$          & 1.48 & 5.23 & $-$60.6 \\
4922\ He\,{\sc i}       & 0.63 & 3.04 & $-$30.7 \\
5876\ He\,{\sc i}       & 1.43 & 4.32 & $-$36.9 \\
6563\ H$\alpha$         & 0.09 (0.49) &  & $-$220.3 \\
6563\ H$\alpha$         & $-$0.86 ($-$0.26) &  & $-$7.7 \\
6678\ He\,{\sc i}       & 0.81 & 3.95 & $-$36.1 \\
\hline
\end{tabular}
}
\end{table}

\subsection{Spectral classification of TYC\,3159-6-1}
\label{sec:clas}

\begin{figure*}
\includegraphics[width=12cm,angle=90]{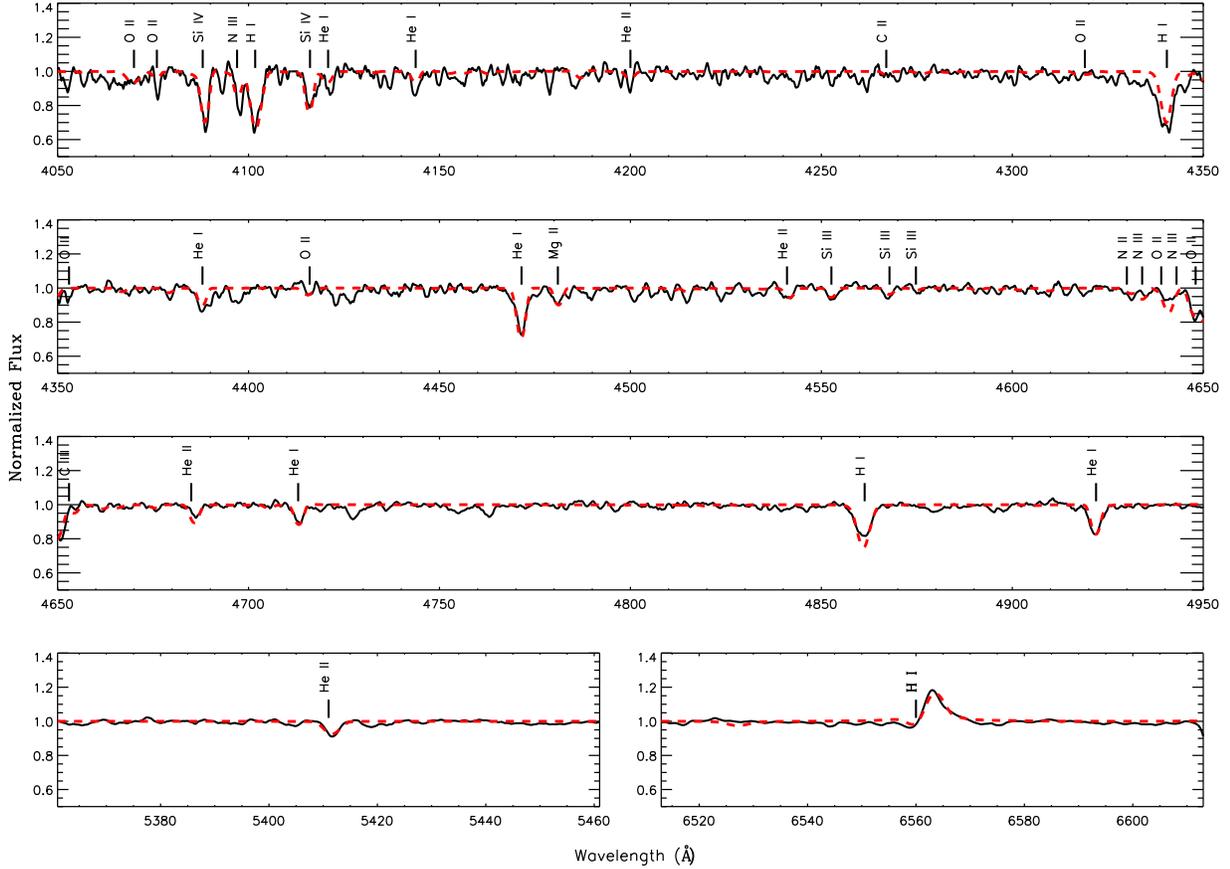}
\centering \caption{Normalized and re-binned spectrum of
TYC\,3159-6-1 taken on 2012 November 14, compared with the
best-fitting {\sc fastwind} model (red dashed line) with the
parameters as given in Table\,\ref{tab:par}. The lines fitted by
the model are highlighted.} \label{fig:spec}
\end{figure*}

Fig.\,\ref{fig:spec} presents the normalized spectrum of
TYC\,3159-6-1 taken on 2012 November 14. The spectrum is dominated
by absorption lines of H\,{\sc i} and He\,{\sc i}. H$\alpha$ is in
emission and shows a P\,Cygni profile. The He\,{\sc ii}
$\lambda\lambda$4200, 4542, 4686 and 5411 absorption lines are
weak, which implies that TYC\,3159-6-1 is of spectral type earlier
than B1 (Walborn \& Fitzpatrick 1990). The Mg\,{\sc ii}
$\lambda$4481 line is weak, which is typical of late O-B0 stars.
The Si\,{\sc iv} $\lambda\lambda$4089, 4116, Si\,{\sc iii}
$\lambda$4552, N\,{\sc iii} $\lambda$4097, and the blend of the
C\,{\sc iii} $\lambda\lambda$4647, 4650, 4651 and O\,{\sc iii}
$\lambda$4650 lines are clearly visible. The high interstellar
extinction towards the star ($\approx 6$ mag; see
Section\,\ref{sec:red}) is manifested in numerous diffuse
interstellar bands (DIBs), of which the most prominent are at
4429, 4727, 4762, 5780, 5797, 5850 and 6614 \AA.

\begin{figure}
\includegraphics[width=8.5cm]{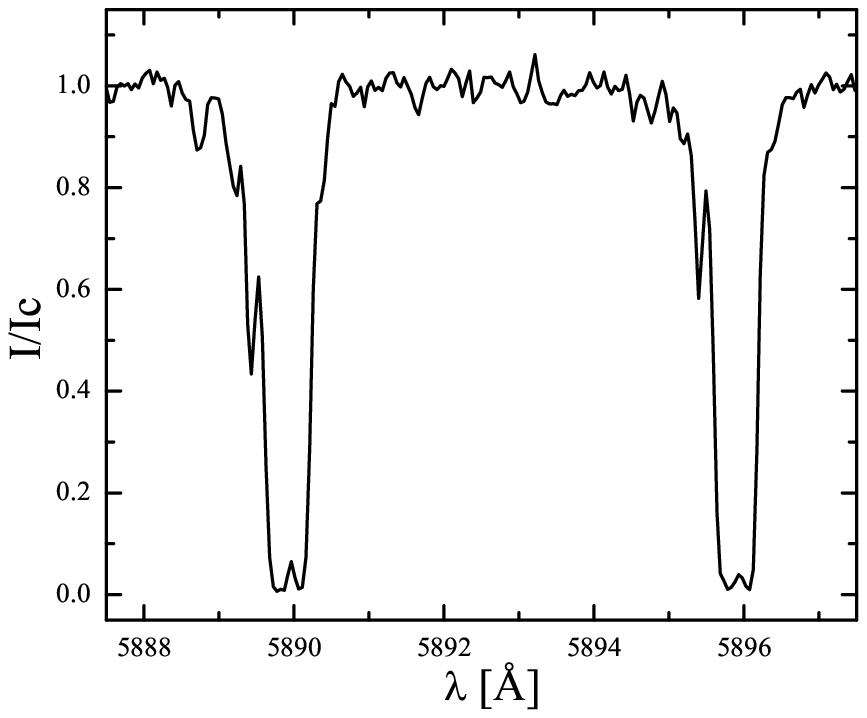}
\centering \caption{Na\,{\sc i} D-lines in the higher-resolution
spectrum of TYC\,3159-6-1 taken on 2012 December 28. Two saturated
components of the lines are of interstellar origin, while the
third (most blueshifted) one may be formed in the circumstellar
environment. The intensity is normalized to the local continuum,
the wavelength scale is heliocentric.} \label{fig:na}
\end{figure}

The higher resolution of the spectrum taken on 2012 December 28
allowed us to resolve the Na\,{\sc i} $\lambda\lambda$5890, 5896
absorption lines into three components with RV$\approx -38, -22$
and $-4 \, \kms$ (see Fig.\,\ref{fig:na}). Two saturated
components of these lines are of interstellar origin. RV of the
third (most blueshifted) component of $-38 \, \kms$ is close to
$v_{\rm r,hel}$, which suggests that this absorption originates in
the circumstellar material.

The relative strength of the He\,{\sc ii} $\lambda$4542 line with
respect to the He\,{\sc i} $\lambda$4388 and Si\,{\sc iii}
$\lambda$4552 ones implies that TYC\,3159-6-1 is of O9.5$-$O9.7
spectral type, while that of the lines He\,{\sc ii} $\lambda4686$
and He\,{\sc i} $\lambda$4713 indicates luminosity class Ib (Sota
et al. 2011). This spectral classification is consistent with the
non-detection of the Si\,{\sc iii} $\lambda$5740 line, which is
observed in the spectra of supergiants of spectral types B0 and
later (Miroshnichenko et al. 2004), but absent in those of
O9$-$O9.5 ones.

\begin{figure}
\includegraphics[width=8.5cm]{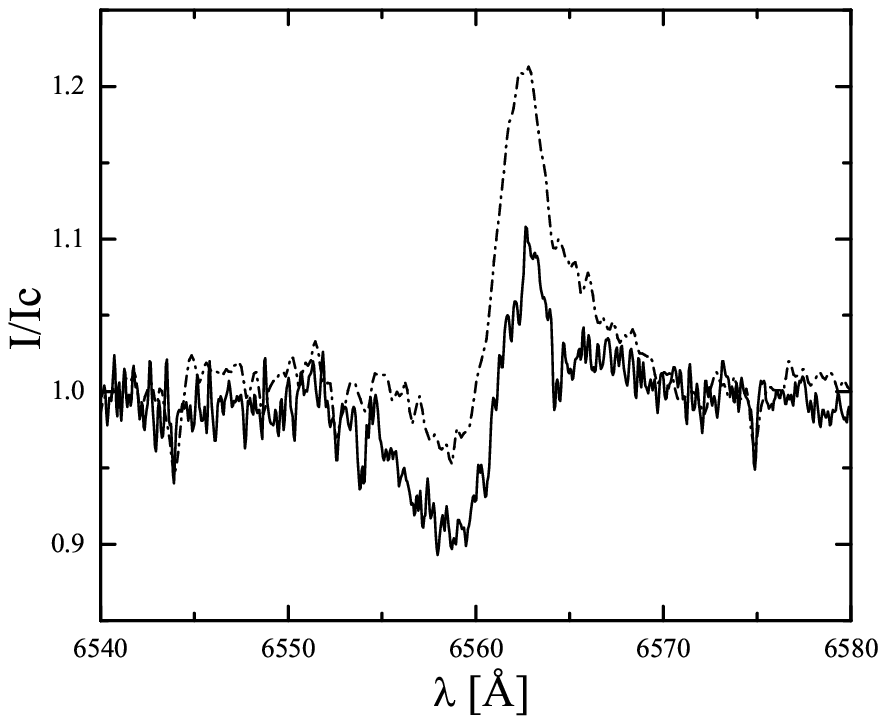}
\centering \caption{H$\alpha$ line profiles in the spectra of
TYC\,3159-6-1. The dash-dotted line shows the profile from 2012
November 14, while the solid line shows the one from 2012 December
28. Weak lines inside and around H$\alpha$ are telluric. The
intensity and wavelength are in the same units as in
Fig.\,\ref{fig:na}.} \label{fig:ha}
\end{figure}

Comparison of the two spectra revealed significant variability in
the H$\alpha$ line profile and the strength of its emission and
absorption components (see Fig.\,\ref{fig:ha} and
Table\,\ref{tab:lines}). Moreover, the emission component in both
spectra became much weaker than that in the spectrum taken five
years earlier by Takita et al. (2009). Such changes in the
H$\alpha$ line (caused by the variability of the stellar wind) are
typical of O supergiants (e.g. Markova et al. 2005).

\subsection{Spectral analysis and stellar parameters}
\label{sec:mod}

To derive the fundamental parameters of TYC\,3159-6-1, we modelled
its spectrum (taken on 2012 November 14) using the stellar
atmosphere code {\sc fastwind} (Fast Analysis of STellar
atmospheres with WINDs; Santolaya-Rey, Puls \& Herrero 1997; Puls
et al. 2005). The code takes into account non-local
thermodynamical equilibrium effects in spherical symmetry with an
explicit treatment of the stellar wind effects by considering a
$\beta$-like wind velocity law (Schaerer \& Schmutz 1994). {\sc
fastwind} generates realistic models in a short period of time
when compared with other similar codes, which is crucial for
building large sets of models.

Characterization of optical spectra of late O- and early B-type
stars can be established on well known lines between $\sim
4000-7000$ \AA \, (e.g. Crowther, Lennon \& Walborn 2006). For
instance, the determination of temperature and surface gravity can
be achieved through the simultaneous fit of the ionization balance
of different ionization stages of the same element (e.g. Si\,{\sc
iv}/Si\,{\sc iii}), and the fit to Balmer line wings. Nonetheless,
the low SNR of the available spectroscopic material makes this
procedure challenging. To tackle this issue, the technique
suggested by Urbaneja et al. (2005) for low SNR spectral analysis
was employed. We selected the main lines observed in
TYC\,3159-6-1, i.e. H\,{\sc i}, He\,{\sc i--ii} and Si\,{\sc
iii--iv}, and looked for the combination of stellar parameters
which best reproduce the spectrum. This analysis was based on a
large grid of {\sc fastwind} stellar atmosphere models and a
package of routines designed for automatic and fast search for
best possible matches through a $\chi^2$ minimization (see Lefever
2007). Afterwards, a set of tailored models was computed by
varying the abundances of the CNO elements in steps of
$0.20\,$dex. A complete description of the technique and the
stellar parameters covered by the grid of stellar models can be
found in Castro et al. (2012).

Each {\sc fastwind} model is defined by nine parameters: effective
temperature, T$_{\rm{eff}}$, surface gravity, $g$, radius,
$R_\ast$, helium abundance, He/H, microturbulence, $\xi$,
metallicity, $Z$, wind velocity law, $\beta$, wind terminal
velocity, $v_\infty$, and mass-loss rate, $\dot{M}$. Some of these
parameters can be constrained using empirical calibrations. The
stellar radius was estimated for the given pair $($T$_{\rm{eff}},
\log g)$ through the flux-weighted gravity--luminosity
relationship (Kudritzki, Bresolin \& Przybilla 2003):
\begin{equation}
M_{\rm bol}=(3.41\pm0.16)(\log g_{\rm F} -1.5)-(8.02\pm0.04) \, ,
\label{eqn:mbol}
\end{equation}
where $M_{\rm bol}$ is the bolometric magnitude and $\log g_{\rm
F} =\log g-4\log (T_{\rm eff}\times10^{-4})$. Since $v_\infty$
cannot be constrained from the optical spectrum alone (Puls, Vink
\& Najarro 2008), we derived it from the escape velocity by using
an empirical calibration [see equation\,(2) in Castro et al.
2012]. Moreover, Puls et al. (1996) showed that different
combinations of $\dot{M}$, $R_\ast$ and $v_\infty$ can produce the
same emergent line profiles as long as the optical depth
invariant,
\begin{equation}
Q={\dot{M} \over (R_\ast v_\infty)^{3/2}} \, , \label{eqn:q}
\end{equation}
remains constant (here $\dot{M}$, $R_\ast$ and $v_\infty$ are in
units of $\msun, \rsun$ and $\kms$, respectively). Since $R_\ast$
and $v_\infty$ are not free parameters, variations of $\dot{M}$
are equivalent to variations of $Q$.

The low SNR prevents us from exploring the possible effect of
macroturbulence on broadening of line shapes (Ryans et al. 2002;
Hunter et al. 2008). Assuming that the broadening is due only to
rotation, we derive a projected rotational velocity of $115 \,
\kms$. A similar estimate can be obtained from the relationship
between the FWHM of the He\,{\sc i} $\lambda$4471 line and the
projected rotational velocity, $v\sin i = 41.25 \, {\rm
FWHM}(4471) \, \kms \,$ (Steele, Negueruela \& Clark 1999), which
for FWHM(4471)=2.60 \AA \, (see Table\,\ref{tab:lines}) and after
correction for the instrumental FWHM of 0.3 \AA \, gives $v\sin i
\approx 107 \, \kms$. These estimates should be considered as an
upper limit to the projected stellar rotational velocity because
the macroturbulence could significantly contribute to the line
broadening.

The best-fitting model for TYC\,3159-6-1 is overlayed on the
observed (normalized) spectrum in Fig.\,\ref{fig:spec}, while the
stellar parameters derived from this model are compiled in
Table\,\ref{tab:par} and show a good agreement with those typical
of O9.5\,I$-$B0\,I stars (e.g. Crowther et al. 2006). The errors
quoted in the table were set at 95 per cent of probability
distributions obtained during the quantitative analysis of the
spectrum.

\begin{table}
\caption{Stellar parameters for TYC\,3159-6-1.}
\label{tab:par}
\centering
\begin{tabular}{lr}
\hline
T$_{\rm{eff}}$ (kK) & 27.0$\pm$2.1 \\
$\log g$ (cgs) & 2.8$\pm$0.2 \\
$R_\ast$ ($\rsun$) (adopted) & 32 \\
He/H (by number) & 0.10$\pm$0.02 \\
$\xi$ ($\kms$) & 32$\pm$13 \\
$v_\infty$ ($\kms$) (adopted) & 1200 \\
$\beta$ & 2.2 \\
$v\sin i$ ($\kms$) & 115 \\
$\log Q$ & $-12.7\pm0.2$ \\
\hline
\end{tabular}
\end{table}

\begin{table}
\caption{CNO elemental abundances (by number) in TYC\,3159-6-1.
The solar abundances are from Asplund et al. (2009).}
\label{tab:ab}
\centering
\begin{tabular}{lrr}
\hline $\log(X/{\rm H})+12$ & TYC\,3159-6-1 & Sun \\
\hline C & $8.0^{+0.5} _{-0.3}$ & 8.43 \\
N & $7.6^{+0.2} _{-0.4}$ ($8.0^{+0.2} _{-0.4}$)$^a$ & 7.83 \\
O & $8.5^{+0.4} _{-0.6}$ & 8.69 \\
\hline
\end{tabular}
\begin{itemize}
\item[$^a$]
 \footnotesize{Based only on the N\,{\sc iii} $\lambda$4097 line. See text for details.}
\end{itemize}
\end{table}

Constraining chemical abundances in low SNR data is also a
challenging process. Nonetheless, the automatic routines were
designed for gaining as much as possible the information hidden in
the spectrum. Table\,\ref{tab:ab} gives the CNO chemical
abundances with the errors established at 95 per cent of the
probability distribution found for each element. All three
abundances are consistent within the margins of error with the
solar ones (Asplund et al. 2009). The carbon abundance taken at
face value shows depletion with respect to the solar one, which is
expected for an atmosphere mixed with processed material from
inner layers (see Langer 2012 and references therein). For
nitrogen we derived two abundances. The first one is based on the
simultaneous fit of all the nitrogen lines detected in the
spectrum and the second one (given in Table\,\ref{tab:ab} in
brackets) was derived by fitting only the N\,{\sc iii}
$\lambda$4097 line (cf. Crowther et al. 2006). Both abundances
agree within errors with the solar one, but the second one might
also indicate a moderate enhancement of the nitrogen. We expect
that the discrepancy in the nitrogen abundance estimates will be
resolved with the next generation of {\sc fastwind} nitrogen
atomic models (Rivero Gon\'{z}alez, Puls \& Najarro 2011). The
oxygen abundance was set by several very weak lines, which are
blurred by the noise (see Fig.\,\ref{fig:spec}), so that it should
be considered as an upper limit.

\subsection{Reddening and distance to TYC\,3159-6-1}
\label{sec:red}

To estimate the distance to TYC\,3159-6-1, one can use the
observed photometry of this star and the synthetic $UBVJHK$
photometry of Galactic O stars by Martins \& Plez (2006), which
for an O9.5\,I star gives the $V$- and $K$-band absolute
magnitudes, $M_V$ and $M_K$, and the intrinsic colours $(B-V)_0$
and $(J-K)_0$ of $-$6.34, $-$5.52, $-$0.26 and $-$0.21 mag,
respectively. Using these figures and the $B,V,J$ and $K_{\rm s}$
magnitudes from Table\,\ref{tab:det}, one can estimate the $V$-
and $K$-band extinction towards TYC\,3159-6-1 and the distance
modulus of this star with help of the relationships:
\begin{equation}
A_V =3.1[(B-V)-(B-V)_0] \, , \label{eqn:av}
\end{equation}
\begin{equation}
A_K =0.66[(J-K)-(J-K)_0] \, , \label{eqn:ak}
\end{equation}
\begin{equation}
DM_V=V-M_V -A_V \, , \label{eqn:dmv}
\end{equation}
\begin{equation}
DM_K=K-M_K -A_K \, , \label{eqn:dmk}
\end{equation}
where the standard total-to-selective absorption ratio
$R=A_V/E(B-V)=3.1$ is assumed, and $K=K_{\rm s}+0.04$ mag
(Carpenter 2001). From equations\,(\ref{eqn:av})-(\ref{eqn:dmk}),
it follows that $A_V =5.86\pm0.89$ mag, $A_K =0.69\pm0.02$ mag,
$DM_V=11.36\pm0.89$ mag and $DM_K=10.82\pm0.02$ mag. For the error
calculation, only the errors of the photometry were considered.
The derived distance moduli correspond to distances of
$1.87^{+0.95} _{-0.63}$ kpc and $1.46\pm0.12$ kpc, respectively.

Although the above two distance estimates are consistent with each
other within the margins of error, we note that the distance based
on the optical photometry might be overestimated because of
anomalous reddening towards the star (e.g. caused by destruction
of dust particles responsible for the visible extinction by
stellar winds, interstellar shocks and/or radiation field). From
equations\,(\ref{eqn:dmv}) and (\ref{eqn:dmk}) it follows that
$DM_V$ and $DM_K$ can be reconciled with each other if
$R\approx3.4$. In this connection, we note that $R=3.4\pm0.1$ was
derived towards the cluster NGC\,6910 (located at $\approx 1$
degree to the southeast of TYC\,3159-6-1; see Fig.\,\ref{fig:dol})
on the basis of photoelectric $UBVR$ photometry for 132 cluster
members (Shevchenko, Ibragimov \& Chernysheva 1991). In what
follows, we adopt the distance based on the 2MASS photometry, i.e.
$d\approx 1.5$ kpc (see also Section\,\ref{sec:par}). At this
distance, 1 arcmin corresponds to $\approx 0.43$ pc.

The distance to TYC\,3159-6-1 can also be estimated using EWs of
the Na\,{\sc i} D-lines. According to Beals \& Oke (1953), $d=1.6
D$ kpc, where $D$ is the average EW of the D$_1$ and D$_2$ lines
in \AA. With EW(D$_1$)=0.81 \AA \, and EW(D$_2$)=0.71 \AA,
measured in the higher resolution (McDonald) spectrum, one obtains
$d\approx 1.2$ kpc. This distance estimate is supported by a
finding by Hobbs (1974) that the EW of the D$_2$ line grows on
average at 0.60 \AA/kpc, which implies $d\approx1.2$ kpc as well
(cf. van Kerkwijk, van Oijen \& van den Heuvel 1989). The
estimates based on the sodium lines should be considered as lower
limits because both lines in the spectrum of TYC\,3159-6-1 are
saturated.

\begin{table}
\caption{Estimates of the colour excess, $E(B-V)$, towards
TYC\,3159-6-1 based on the EWs of DIBs at $\lambda 5780$, $\lambda
5797$ and $\lambda$8620.} \label{tab:dib}
\begin{center}
\begin{tabular}{cccc}
\hline
DIB & EW($\lambda$) & $E(B-V)$ \\
(\AA) & (\AA) & (mag) \\
\hline
5780 & 0.68$\pm$0.03 & 1.34$\pm$0.10  \\
5797 & 0.26$\pm$0.03 & 1.73$\pm$0.10  \\
8620 & 0.43$\pm$0.03 & 1.16$\pm$0.09  \\
\hline
\end{tabular}
\end{center}
\end{table}
\begin{table*}
\begin{minipage}{0.72\textwidth}
\caption{Proper-motion, heliocentric radial velocity, peculiar
transverse (in Galactic coordinates) and radial velocities, and
the total space velocity of TYC\,3159-6-1.}
\label{tab:prop}
\begin{tabular}{@{}lccccccccccccccc@{}}
\hline $\mu _\alpha \cos \delta$ & $\mu _\delta$ & $v_{\rm r, hel}$ & $v_{\rm l}$ & $v_{\rm b}$ & $v_{\rm r}$ & $v_\ast$ \\
(mas ${\rm yr}^{-1}$) & (mas ${\rm yr}^{-1}$) & ($\kms$) & ($\kms$) & ($\kms$) & ($\kms$) & ($\kms$) \\
\hline $-2.4\pm1.0$ & $-0.1\pm1.1$ & $-35.8\pm3.0$ & $25.3\pm7.4$ & $20.1\pm7.1$ & $-25.8\pm5.0$ & $41.3\pm6.5$ \\
\hline
\end{tabular}\\
\end{minipage}
\end{table*}

Alternatively, the extinction (and distance) towards TYC\,3159-6-1
can be derived using the correlation between the intensity of the
DIBs and the colour excess $E(B-V)$ (see Herbig 1995 for a
review). In Table\,\ref{tab:dib} we provide estimates of $E(B-V)$
based on EWs of DIBs at $\lambda 5780$, $\lambda 5797$ and
$\lambda$8620\footnote{The latter DIB was detected in the McDonald
spectrum of TYC\,3159-6-1.}, and the relationships given in Herbig
(1993) and Munari et al. (2008). Using these estimates and
assuming $R=3.4$, from equation\,(\ref{eqn:dmv}) one can obtain a
distance towards TYC\,3159-6-1 of $\approx 3.4, 1.9$ and 4.5 kpc,
respectively. The discrepancy between these distance estimates and
that based on the 2MASS photometry could be due to a patchy
distribution of the obscuring material (and the carriers of the
DIBs) in the direction of Cygnus\,X (Schneider et al. 2006).

\subsection{Luminosity and evolutionary status of TYC\,3159-6-1}
\label{sec:pro}

Using equation\,(\ref{eqn:mbol}) and $\log g$ and $T_{\rm eff}$
from Table\,\ref{tab:par}, one finds the bolometric magnitude of
TYC\,3159-6-1 of $M_{\rm bol}=-9.47\pm0.83$ mag, which translates
into the bolometric luminosity of $\log(L/{\rm
L}_\odot)=0.4(4.74-M_{\rm bol})=5.68\pm0.33$. This luminosity
implies that the zero-age main-sequence (ZAMS) mass of
TYC\,3159-6-1 was of $40\pm15 \, \msun$ (e.g. Ekstr\"{o}m et al.
2012). Then using $R_\ast$ and $\log g$, one can derive the
current mass of TYC\,3159-6-1 of $M_\ast \approx 23^{+14} _{-9} \,
\msun$ (the quoted errors are based only on the errors of the
gravity and did not take into account the possible range of
$R_\ast$ implied by uncertainties in the luminosity and
temperature). Taken at face value, this mass implies that
TYC\,3159-6-1 has already lost about a half of its initial mass
and now experiences a blue loop after the brief red supergiant
phase of evolution (e.g. Ekstr\"{o}m et al. 2012). This
possibility however is inconsistent with the He and CNO abundances
derived for TYC\,3159-6-1, which rather suggest that the star only
recently evolved off the main sequence and therefore preserved
most of its ZAMS mass.

To further constrain $M_\ast$ and thereby to estimate the age of
TYC\,3159-6-1, we compare its He and CNO abundances (see
Tables\,\ref{tab:par} and \ref{tab:ab} in Section\,\ref{sec:mod})
with those predicted by stellar evolutionary models (e.g., Brott
et al. 2011; Ekstr\"{o}m et al. 2012). Using the grid of models by
Brott et al. (2011), we found that the main stellar parameters of
TYC\,3159-6-1 (temperature, luminosity, radius, gravity and
abundances) can be matched very well with the $40 \, \msun$ model,
provided that the age of the star is $\approx 4.0-4.5$ Myr.
Particularly, a $40 \, \msun$ star with the initial rotation
velocity of $161 \, \kms$ would have at the age of $\approx 4$ Myr
the following parameters: $T_{\rm eff}\approx 25.7$ kK,
$log(L/\lsun)\approx5.62$, $R_* \approx 32.6 \, \rsun$, $\log
g\approx2.97$, $v\sin i \approx 106 \, \kms$. The CNO abundances
of this star are 8.06, 7.88 and 8.53, respectively. All these
parameters and abundances are in good agreement with those derived
for TYC\,3159-6-1 (see Tables\,\ref{tab:par} and \ref{tab:ab}).
The Brott et al.'s models also suggest that the current mass of
TYC\,3159-6-1 should be $\approx 36 \, \msun$, which, considering
the uncertainties on the stellar parameters, is consistent with
the estimate of $M_\ast$ given above. Thus, we conclude that
TYC\,3159-6-1 is a redward evolving star, which has only recently
entered into the blue supergiant phase.

\section{Discussion}
\label{sec:dis}

\subsection{TYC\,3159-6-1 as a runaway}
\label{sec:run}

The enhanced brightness of the nebula along its northeast rim
could be caused by motion of TYC\,3159-6-1 in the northeast
direction (cf. Danforth \& Chu 2001; Gvaramadze et al. 2009). To
check this possibility, we searched for proper motion measurements
for TYC\,3159-6-1 using the VizieR catalogue access
tool\footnote{http://vizier.u-strasbg.fr/cgi-bin/VizieR}. We found
several measurements of which the most recent one (and the one
with the smallest claimed errors) is provided by the fourth U.S.
Naval Observatory CCD Astrograph Catalog (UCAC4; Zacharias et al.
2013). This measurement is given in Table\,\ref{tab:prop} along
with the heliocentric radial velocity of the star (derived in
Section\,\ref{sec:obs}), the components of its peculiar transverse
velocity (in Galactic coordinates), $v_{\rm l}$ and $v_{\rm b}$,
the peculiar radial velocity, $v_{\rm r}$, and the total peculiar
(space) velocity, $v_\ast$. To derive the peculiar velocities, we
used the Galactic constants $R_0 = 8.0$ kpc and $\Theta _0 =240 \,
\kms$ (Reid et al. 2009) and the solar peculiar motion
$(U_{\odot},V_{\odot},W_{\odot})=(11.1,12.2,7.3) \, \kms$
(Sch\"onrich, Binney \& Dehnen 2010). For the error calculation,
only the errors of the proper motion and the radial velocity
measurements were considered.

The derived space velocity of $\approx40 \, \kms$ implies that
TYC\,3159-6-1 is a classical runaway star (Blaauw 1961;
Cruz-Gonz\'{a}lez et al. 1974). From Table\,\ref{tab:prop} it
follows that TYC\,3159-6-1 is moving in the northeast direction,
i.e. towards the brightest arc of the IR nebula, with a transverse
velocity of $\approx 32 \, \kms$. Moreover, TYC\,3159-6-1 is
approaching us with a velocity of $\approx 25 \, \kms$, so that
the vector of its space velocity makes an angle of $\approx 50$
degree with respect to our line-of-sight.

\begin{figure}
\includegraphics[width=8cm]{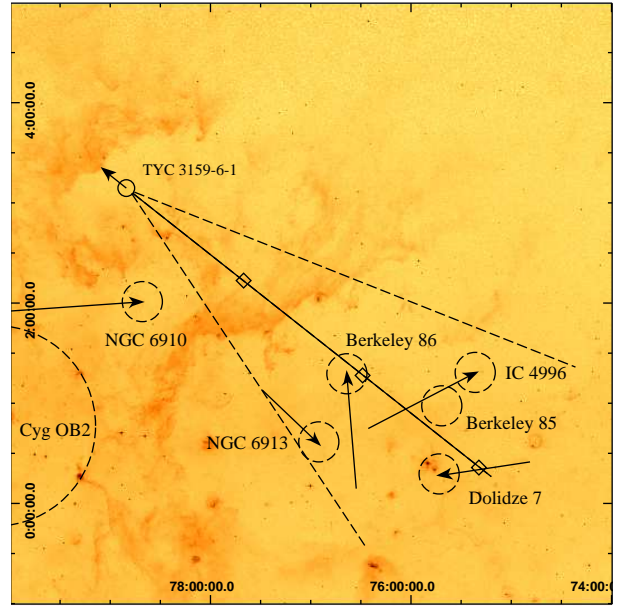}
\centering \caption{$6\degr \times 6\degr$ {\it MSX} 8.3\,$\mu$m
image of the Cygnus X region centred at $l=77\degr, b=2\degr$,
with the position of TYC\,3159-6-1 indicated by a circle. The
arrow shows the direction of motion of TYC\,3159-6-1, while a
solid line indicates the trajectory of TYC\,3159-6-1 (with 1 sigma
uncertainties shown by dashed lines). The positions of the star 1,
2 and 3 Myr ago are marked by diamonds. The positions of six star
clusters located near to or within the error cone of the past
trajectory of TYC3159-6-1 are indicated by small dashed circles.
The arrows attached to these circles show the direction of
clusters' peculiar motions, while their origins correspond to the
positions of the clusters 4 Myr ago. The approximate boundary of
the Cyg\,OB2 association is shown by a dashed circle of a diameter
of 2 degree. The image is oriented with Galactic longitude (in
units of degrees) increasing to the left and Galactic latitude
increasing upwards. At a distance of 1.6 kpc, 1 degree corresponds
to $\approx27.5$ pc. See text for details.} \label{fig:dol}
\end{figure}

\subsection{Parent cluster to TYC\,3159-6-1}
\label{sec:par}

\begin{table}
\caption{Distance and age estimates for star clusters located
within or close to the error cone of the past trajectory of
TYC\,3159-6-1.} \label{tab:clu}
\begin{center}
\begin{tabular}{cccc}
\hline
Object & $d$ & age \\
 & (kpc) & (Myr) & \\
\hline
IC\,4996 & 1.6$^{a,b}$; 1.7$^c$; 2.0$^d$; 2.3$^e$ & 6.3$^c$; 9$^a$; 10$^b$ \\
Berkeley\,85 & 1.6$^d$; 1.8$^b$ & 1000$^b$ \\
Berkeley\,86 & 1.6$^e$; 1.7$^{f,d}$; 1.9$^g$ & 2$-$3$^g$; 6$^h$; 8.7$^f$ \\
Dolidze\,7 & 1.2$^i$; 1.5$^e$; 1.6$^j$; 1.9$^d$ & 3$^j$; 4$^k$; 4$-$6$^l$ \\
NGC\,6910 & 1.6$^m$; 1.8$^d$; 2.3$^e$ & 6$^m$; 6.5$^e$ \\
NGC\,6913 & 1.1$^n$; 1.8$^d$; 2.2$^g$ & 4$-$6$^g$\\
\hline
\end{tabular}
\end{center}
$^{a}$Vancevicius et al. (1996); $^{b}$Maciejewski \& Niedzielski
(2007); $^c$Kharchenko et al. (2005); $^d$Le Duigou \&
Kn\"{o}dlseder (2002); $^e$Bhavya et al. (2007); $^f$Delgado,
Alfaro \& Cabrera-Cano (1997); $^g$Massey, Johnson \&
Degioia-Eastwood (1995); $^h$Forbes (1981); $^i$Turner et al.
(2006); $^j$Massey, DeGioia-Eastwood \& Waterhouse (2001);
$^k$Polcaro \& Norci 1998; $^l$Oskinova et al. 2010;
$^m$Kolaczkowski et al. (2004); $^n$Wang \& Hu (2000).
\end{table}
\begin{table*}
\begin{minipage}{0.6\textwidth}
\caption{Proper-motion measurements for star clusters located near
to or within the error cone of the past trajectory of
TYC\,3159-6-1. For each measurement, the components of the
peculiar transverse velocity (in Galactic coordinates) and the
total space velocity, $v_{\rm tot}$, are calculated and added to
the table. See text for details.} \label{tab:pro}
\begin{tabular}{@{}lccccccccccccccc@{}}
\hline Object & $\mu _\alpha \cos \delta$ & $\mu _\delta$ & $v_{\rm l}$ & $v_{\rm b}$ & $v_{\rm tot}$ \\
& (mas ${\rm yr}^{-1}$) & (mas ${\rm yr}^{-1}$) & ($\kms$) & ($\kms$) & ($\kms$) & \\
\hline
IC\,4996 & $-3.2\pm0.1$ & $-5.5\pm0.1$ & $-7.7\pm0.8$ & $4.0\pm0.8$ & $8.7\pm0.8$ \\
Berkeley\,86 & $-3.1\pm0.3$ & $-4.2\pm0.2$ & $0.7\pm1.8$ & $8.5\pm2.1$ & $8.5\pm2.1$ \\
Dolidze\,7 & $-1.6\pm0.7$ & $-4.2\pm0.5$ & $6.6\pm4.1$ & $-1.0\pm4.6$ & $6.7\pm4.1$ \\
NGC\,6910 & $-3.4\pm0.3$ & $-6.2\pm0.3$ & $-13.6\pm1.9$ & $1.2\pm1.9$ & $13.6\pm1.9$ \\
NGC\,6913 & $-1.5\pm0.4$ & $-4.7\pm0.3$ & $-4.2\pm2.7$ & $-4.0\pm2.7$ & $5.8\pm2.7$ \\
\hline
\end{tabular}\\
\end{minipage}
\end{table*}

Fig.\,\ref{fig:dol} shows the {\it Midcourse Space Experiment}
({\it MSX}) satellite (Price et al. 2001) $6\degr \times 6\degr$
image of the Cygnus X region. The position of TYC\,3159-6-1 and
the direction of its peculiar transverse velocity are indicated by
a circle and an arrow, respectively. The solid line represents the
trajectory of TYC\,3159-6-1, as suggested by the proper motion
measurements, while the dashed ones indicate 1 sigma
uncertainties. Fig.\,\ref{fig:dol} also shows positions of six
star clusters, five of which are located within the error cone of
the past trajectory of TYC\,3159-6-1. The remaining one,
NGC\,6910, is located not far from the cone and potentially can be
the parent cluster to TYC\,3159-6-1, provided that its space
velocity is pointed away from the cone. Let us discuss which of
these six clusters might be the birthplace of TYC\,3159-6-1.

In Table\,\ref{tab:clu} we summarize distance and age estimates
for the six clusters (listed according to their right ascensions)
as derived from literature. The very old age of Berkeley\,85
allows us to exclude this cluster from further consideration. From
the table it follows that the main problem with ascertaining the
parent cluster to TYC\,3159-6-1 is the uncertainty in distances to
the clusters, which are based on spectrophotometric studies. This
uncertainty is mainly caused by high and patchy extinction towards
the Cygnus\,X region (e.g. Schneider et al. 2006) and by
variations of reddening across the face of particular clusters
(e.g. Turner \& Forbes 1982; Wang \& Hu 2000). On the other hand,
there are strong indications that most molecular clouds in
Cygnus\,X, and correspondingly their associated star clusters,
form a coherent complex (Schneider et al. 2006; cf. Mel'nik \&
Efremov 1995), which is located at about the same distance as the
Cyg\,OB2 association (i.e. at $d\approx1.4$ kpc; Rygl et al.
2012). The range of distances derived for the clusters in
Table\,\ref{tab:clu} allows the possibility that all of them might
be the birthplace of TYC\,3159-6-1. In the following we adopt a
common distance of 1.6 kpc to all these clusters, where we took
into account the fact that TYC\,3159-6-1 is moving towards us with
the velocity of $\approx25 \, \kms$ and therefore might travel
$\approx100$ pc in the radial direction during its lifetime. We
use this distance and the cluster proper motion measurements by
Loktin \& Bechenov (2003) to calculate the magnitude and direction
of space velocities of the clusters and thereby to check whether
their trajectories intersect in the past with that of
TYC\,3159-6-1 (cf. Schilbach \& R\"{o}ser 2008; Gvaramadze \&
Bomans 2008). The results of the calculations are summarized in
Table\,\ref{tab:pro}. The derived space velocities of the clusters
of $\sim 10 \, \kms$ are comparable to those of several
star-forming regions around the Cyg\,OB2 association (Rygl et al.
2012). It is worth noting that these velocities do not show a
regular pattern, which suggests the presence of large-scale
turbulent motions in the parent molecular cloud of the clusters.

In Fig.\,\ref{fig:dol} we show the trajectories of the clusters
and their positions on the sky 4 Myr ago. One can see that
NGC\,6910 is moving in the ``incorrect" direction, so that it
cannot be the parent cluster to TYC\,3159-6-1. The trajectories of
the remaining four clusters are confined within the error cone of
the past trajectory of TYC3159-6-1 and therefore, in principle,
they can intersect the trajectory of the star. Note however that
the age estimates for IC\,4996 (6$-$10 Myr) are inconsistent with
the age of TYC\,3159-6-1 of 4 Myr, which suggests that the cluster
and the star are not related to each other. Moreover, although we
cannot exclude the possibility that TYC\,3159-6-1 was ejected from
Berkeley\,86 or NGC\,6913, we note that these two clusters were at
the periphery of the cone $\sim 3$ Myr ago, which makes Dolidze\,7
the more likely candidate for the birthplace of the star. An
indirect support to the physical relationship between
TYC\,3159-6-1 and Dolidze\,7 comes from the presence in this
cluster of the candidate LBV star V439\,Cyg (Polcaro \& Norci
1998; Polcaro, Norci \& Miroshnichenko 2006), whose age and ZAMS
mass of 4 Myr and $40 \, \msun$, respectively (Polcaro \& Norci
1998), are similar to those of TYC\,3159-6-1.

It is also worth noting that the kinematic age of TYC\,3159-6-1,
i.e. the time elapsed since the ejection event, constitutes a
significant fraction of its evolutionary one. This implies that
the ejection of TYC\,3159-6-1 into the field cannot be caused by a
supernova explosion in a massive binary system (the companion star
would simply have no time to end its life in a supernova), but is
the result of dynamical three- or four-body encounter in the core
of the parent cluster (cf. Gualandris, Portegies Zwart \& Eggleton
2004; Gvaramadze et al. 2011).

\subsection{Origin of the nebula around TYC\,3159-6-1}
\label{sec:ori}

The orientation of the peculiar transverse velocity of
TYC\,3159-6-1 and the arc-like appearance of the brightest part of
the IR nebula (see the {\it WISE} 8\,$\mu$m image in
Fig.\,\ref{fig:neb}) suggest that the nebula might be a bow shock
created because of interaction between the wind of the
supersonically moving star and the ambient ISM. In this case, the
radius of the arc would correspond to the stand-off distance of
the bow shock, $R_0$, which can be expressed through $\dot{M}$,
$v_\infty$, $v_\ast$, and the number density of the local ISM,
$n_0$, as follows:
\begin{equation}
\label{eqn:bow} R_0 =\left({\dot{M} v_\infty \over 4\pi\rho v_\ast
^2}\right)^{1/2} \, ,
\end{equation}
where $\rho =\mu m_H n_0$, $\mu =1.4$ is the mean molecular
weight, and $m_{\rm H}$ is the mass of a hydrogen atom. For $R_0
\approx0.46$ pc and $\dot{M}\approx 1.5\times10^{-6} \, \msun \,
{\rm yr}^{-1}$, derived from equation\,(\ref{eqn:q}), one finds
from equation\,(\ref{eqn:bow}) that $n_0 \approx 12 \, {\rm
cm}^{-3}$, i.e. a quite reasonable figure (cf. Gvaramadze \&
Bomans 2008). The bow shock interpretation for the nebula around
TYC\,3159-6-1, however, does not allow us to explain the complex
appearance of the nebula as a whole, especially the northern
curved filaments, which might be a signature of a bipolar outflow
from the star. Instead, we suggest that the nebula is composed of
material ejected by TYC\,3159-6-1 in several successive episodes
of enhanced wind mass loss or outbursts, similar to those observed
in LBVs. Our suggestion is supported by the detection of the
circumstellar component in the sodium doublet
(Section\,\ref{sec:clas}), indicating the presence of significant
amount of matter comoving with the star.

Several examples of bipolar nebulae are known to be associated
with blue supergiants, of which the most spectacular are a
triple-ring nebula produced by the progenitor star of the
SN\,1987A (Burrows et al. 1995) and hourglass-shaped nebulae
around the Galactic blue supergiants Sher\,25 (Brandner et al.
1997) and MN18 (Gvaramadze et al. 2010). Similar bipolar nebulae
are also observed around candidate LBVs HD\,168625 (Smith 2007)
and MN13 (Gvaramadze et al. 2010; Wachter et al. 2011). The origin
of these nebulae is still subject of debate, but there is
consensus that different mechanisms might be responsible for their
ejection and shaping (e.g. Morris 1981; Garc\'{i}a-Segura, Langer
\& Mac Low 1996; Chita et al. 2008).

Although all known (bipolar) circumstellar nebulae around massive
stars are associated with stars evolved off the main sequence, it
is still not clear at what post-main-sequence evolutionary phase
they are ejected. According to most scenarios, this should happen
when the star has already undergone a red supergiant phase, i.e.
when the star is close to the endpoint of its life -- the type II
supernova explosion. Detection of the triple-ring nebula around
the SN\,1987A confirms that at least some circumstellar nebulae
originate during the final phases of stellar evolution. On the
other hand, analysis of chemical abundances in some nebulae
associated with massive stars (including the nebula around
Sher\,25) argues against the possibility that these stars have
evolved through the red supergiant phase, and suggests that they
were ejected soon after the end of the main-sequence, i.e. at the
beginning of the blue supergiant phase (Lamers et al. 2001; Hendry
et al. 2008). The mildly enhanced He and N abundances derived for
these nebulae can be understood if by the moment of ejection the
stellar envelopes were enriched by processed material from the
core because of rotationally induced mixing at or near the end of
the main sequence. Accordingly, the central stars of these nebulae
should be fast rotators for the mixing to occur.

As discussed in Section\,\ref{sec:pro}, the He and CNO abundances
in the photosphere of TYC\,3159-6-1 imply that the star only
recently entered into the blue supergiant phase, which in turn
implies that the nebula was ejected just after the main sequence.
It is likely that the origin of this and other nebulae associated
with (unevolved) blue supergiants is related to the fast rotation
of their central stars (Lamers et al. 2001; Hendry et al. 2008).
The post-main-sequence expansion of a fast-rotating star may
result in the nearly critical rotation of its surface layers
(Eriguchi et al. 1992), i.e. may bring the stellar envelope close
to the $\Omega$ limit (Langer 1997, 1998; Maeder \& Meynet 2000),
which in turn may result in enhanced stationary or eruptive mass
loss. Note that the moderate rotational velocity of TYC\,3159-6-1
(see Section\,\ref{sec:mod}) does not contradict to the
possibility that this star was a fast rotator in the near past
because its envelope might already lost a significant fraction of
its angular momentum due to several episodes of mass ejection.

The morphology of the IR nebula associated with TYC\,3159-6-1
suggests that it might be produced in three successive episodes of
enhanced mass loss (outbursts) alternated with periods of
quiescent mass loss. We speculate that the material ejected during
the outbursts is concentrated near the equatorial plane of the
fast-rotating star and that the ram pressure of the ISM due to the
stellar motion makes this material lagging behind the star.
Assuming that the ejected material expands with a velocity of
$\sim 50 \, \kms$ (which is typical of LBVs and related objects;
Nota et al. 1995), one finds the dynamical age of the nebula of
$\sim 3\times10^4$ yr. Moreover, from the almost equal separation
between the southern edge of the nebula and the two concentric
filaments within it one can derive that the duty cycle of the
eruptive activity is of the order of $\sim 10^4$ yr. On the other
hand, the presence of the curved filaments to the north of
TYC\,3159-6-1 suggests that the ejected material has a bipolar
component as well. The polar outflows might originate
simultaneously with the equatorial one (i.e. during the outburst)
or because of collimation of the quiescent stellar wind by the
dense material of the equatorial ejecta.

Assuming that the nebula is mainly composed of the material
ejected during three successive outbursts of equal duration,
$\tau$, and that $\dot{M}$ during these outbursts was $100\times$
the current mass-loss rate (e.g. Humphreys \& Davidson 1994), one
finds $\tau\sim 100$ yr, where we used the mass of the nebula of
$\sim 0.06 \, \msun$ (see Section\,\ref{sec:neb}). Similarly,
using the dynamical age of the nebula and the current $\dot{M}$ of
TYC\,3159-6-1, one finds that the star has lost $\approx 0.05 \,
\msun$ during the quiescent phases, which is comparable to the
mass of the nebula. It is likely that the quiescent
(high-velocity) wind escapes in the polar directions of the nebula
and that it may be responsible for the origin of the curved
filaments to the north of the nebula.

To conclude, we note that virtually in all scenarios for the
origin of circumstellar nebulae around evolved massive stars it is
assumed that these stars are static and therefore are surrounded
by extended low-density wind bubbles formed during the main
sequence. Correspondingly, it is assumed that the structure of
these nebulae is not affected by density inhomogeneities in the
ISM. In reality, however, the majority of known LBVs and related
stars are located outside of any known star cluster and therefore
most likely are runaways (Gvaramadze et al. 2012b). From this it
follows that the field post-main-sequence massive stars are
surrounded by the almost pristine ISM (whose structure is affected
only by the stellar ionizing emission) and therefore their nebulae
may directly interact with this (inhomogeneous) medium. The
optical filaments to the south of TYC\,3159-6-1 (see
Fig.\,\ref{fig:neb}) might be a signature of such an interaction.

\section{Acknowledgements}

This work was partially supported by the DGAPA/PAPIIT project
IN-103912. A.S.M. acknowledges financial support from the
University of North Carolina at Greensboro and from its Department
of Physics and Astronomy. We are grateful to the referee for
careful reading the manuscript and useful suggestions. This work
has made use of the NASA/IPAC Infrared Science Archive, which is
operated by the Jet Propulsion Laboratory, California Institute of
Technology, under contract with the National Aeronautics and Space
Administration, the SIMBAD database and the VizieR catalogue
access tool, both operated at CDS, Strasbourg, France.


\begin{thebibliography}{}
%
\bibitem{} Asplund M., Grevesse N., Sauval A.J., Scott P., 2009, ARA\&A, 47, 481
\bibitem{} Beals C.S., Oke J.B., 1953, MNRAS, 113, 530
\bibitem{} Bhavya B., Mathew B., Subramaniam A., 2007, Bull. Astron. Soc. India,
35, 383
\bibitem{} Blaauw A., 1961, Bull. Astron. Inst. Netherlands, 15, 265
\bibitem{} Brandner W., Grebel E.K., Chu Y.-H., Weis K., 1997, ApJ, 475, L45
\bibitem{} Brott I. et al., 2011, A\&A, 530, A115
\bibitem{} Burrows C.J. et al., 1995, ApJ, 452, 680
\bibitem{} Carpenter J.M., 2001, AJ, 121, 2851
\bibitem{} Castro N. et al., 2012, A\&A, 542, A79
\bibitem{} Chita S.M., Langer N., van Marle A.J., Garcia-Segura G., Heger A., 2008, A\&A, 488, L37
\bibitem{} Crowther P.A., Lennon D.J., Walborn N.R., 2006, A\&A, 446, 279
\bibitem{} Cruz-Gonz\'{a}lez C., Recillas-Cruz E., Costero R., Peimbert M., Torres-Peimbert S., 1974, Rev. Mex. Astron. Astrofis., 1, 211
\bibitem{} Cutri R.M. et al., 2003, VizieR Online Data Catalog, 2246, 0
\bibitem{} Cutri R.M. et al., 2012, VizieR Online Data Catalog, 2311, 0
\bibitem{} Danforth C.W., Chu Y.-H., 2001, ApJ, 552, L155
\bibitem{} Delgado A.J., Alfaro E.J., Cabrera-Cano J., 1997, AJ, 113, 713
\bibitem{} Droege T.F., Richmond M.W., Sallman M.P., Creager R.P., 2006, PASP, 118, 1666
\bibitem{} Egret D., Didelon P., Mclean B.J., Russell J.L., Turon C., 1992, A\&A, 258, 217
\bibitem{} Ekstr\"{o}m S. et al., 2012, A\&A, 537, A146
\bibitem{} Eriguchi Y., Yamaoka H., Nomoto K., Hashimoto M., 1992, ApJ, 392, 243
\bibitem{} Fazio G.G. et al., 2004, ApJS, 154, 10
\bibitem{} Forbes D., 1981, PASP, 93, 441
\bibitem{} Garc\'{i}a-Segura G., Langer N., Mac Low M.-M., 1996, A\&A, 305, 229
\bibitem{} Gualandris A., Portegies Zwart S., Eggleton P.P., 2004, MNRAS, 350, 615
\bibitem{} Gvaramadze V.V., Bomans D.J., 2008, A\&A, 490, 1071
\bibitem{} Gvaramadze V.V., Menten K.M., 2012, A\&A, 541, A7
\bibitem{} Gvaramadze V.V., Kniazev A.Y., Fabrika S., 2010, MNRAS, 405, 1047
\bibitem{} Gvaramadze V.V., Kniazev A.Y., Chen\'{e} A.-N., Schnurr O., 2013, MNRAS, 430, L20
\bibitem{} Gvaramadze V.V., Kniazev A.Y., Kroupa P., Oh S., 2011, A\&A, 535, A29
\bibitem{} Gvaramadze V.V., Weidner C., Kroupa P., Pflamm-Altenburg J., 2012b, MNRAS, 424, 3037
\bibitem{} Gvaramadze V.V. et al., 2009, MNRAS, 400, 524
\bibitem{} Gvaramadze V.V. et al., 2012a, MNRAS, 421, 3325
\bibitem{} Hendry M.A., Smartt S.J., Skillman E.D., Evans C.J., Trundle C., Lennon D.J., Crowther P.A., Hunter I., 2008, MNRAS, 388, 1127
\bibitem{} Herbig G.H., 1995, ARA\&A 33, 19
\bibitem{} Herbig G.H., 1993, ApJ, 407, 142
\bibitem{} Hobbs L.M., 1974, ApJ, 191, 381
\bibitem{} H$\o$g E. et al., 2000, A\&A, 355, L27
\bibitem{} Hora J.L. et al., 2008, New Light on Young Stars: Spitzer's View of Circumstellar
Disks (http://www.ipac.caltech.edu/spitzer2008/proceedings.html)
\bibitem{} Humphreys R.M., Davidson K., 1994, PASP, 106, 1025
\bibitem{} Hunter I., Lennon D.J., Dufton P.L., Trundle C., Sim\'{o}n-D\'{i}az S., Smartt S.J., Ryans R.S.I., Evans C.J., 2008, A\&A, 479, 541
\bibitem{} Huthoff F., Kaper L., 2002, A\&A, 383, 999
\bibitem{} Kharchenko N.V., Piskunov A.E., R\"{o}ser S., Schilbach E., Scholz R.-D., 2005, A\&A, 438, 1163
\bibitem{} Kolaczkowski Z., Pigulski A., Kopacki G., Michalska G., 2004, Acta Astron., 54, 33
\bibitem{} Kudritzki R.P., Bresolin F., Przybilla N., 2003, ApJ, 582, L83
\bibitem{} Lamers H.J.G.L.M., Nota A., Panagia N., Smith L.J., Langer N., 2001, ApJ, 551, 764
\bibitem{} Langer N., 1997, in Nota A., Lamers H., eds, ASP Conf. Ser. Vol. 120, Luminous Blue Variables: Massive
Stars in Transition.Astron. Soc. Pac., San Francisco, p. 83
\bibitem{} Langer N., 1998, A\&A 329, 551
\bibitem{} Langer N., 2012, ARA\&A, 50, 107
\bibitem{} Le Duigou J.M., Kn\"{o}dlseder J., 2002, A\&A, 392, 869
\bibitem{} Lefever K., 2007, PhD thesis, K. U. Leuven
\bibitem{} Levine S., Chakrabarty D., 1995, IA-UNAM Technical Report MU-94-04
\bibitem{} Loktin A.V., Beshenov G.V., 2003, Astron. Rep., 47, 6
\bibitem{} Maciejewski G., Niedzielski A., 2007, A\&A, 467, 1065
\bibitem{} Maeder A., Meynet G., 2000, A\&A, 361, 159
\bibitem{} Markova N., Puls J., Scuderi S., Markov H., 2005, A\&A, 440, 1133
\bibitem{} Martins F., Plez B., 2006, A\&A, 457, 637
\bibitem{} Massey P., Johnson K.E., Degioia-Eastwood K., 1995, ApJ, 454, 151
\bibitem{} Massey P., DeGioia-Eastwood K., Waterhouse E., 2001, AJ, 121, 1050
\bibitem{} McLean B.J., Greene G.R., Lattanzi M.G., Pirenne B., 2000, in Manset N., Veillet C., Crabtree D., eds,
ASP Conf. Ser. Vol. 216, Astronomical Data Analysis Software and
Systems IX. Astron. Soc. Pac., San Francisco, p. 145
\bibitem{} Mel'nik A.M., Efremov Y.N., 1995, Astron. Lett., 21, 10
\bibitem{} Miroshnichenko A.S. et al., 2004, A\&A, 417, 731
\bibitem{} Mizuno D.R. et al., 2010, AJ, 139, 1542
\bibitem{} Morris M., 1981, ApJ, 249, 572
\bibitem{} Munari U. et al., 2008, A\&A, 488, 969
\bibitem{} Nota A., Livio M., Clampin M., Schulte-Ladbeck R., 1995, ApJ, 448, 788
\bibitem{} Oskinova L.M., Gruendl R.A., Ignace R., Chu Y.-H., Hamann W.-R., Feldmeier A., 2010, ApJ, 712, 763
\bibitem{} Peri C.S., Benaglia P., Brookes D.P., Stevens I.R., Isequilla N.L., 2012, A\&A, 538, A108
\bibitem{} Polcaro V.F., Norci L., 1998, A\&A, 339, 75
\bibitem{} Polcaro V.F., Norci L., Miroshnichenko A.S., 2006, in Kraus M., Miroshnichenko A.S., eds, Stars with the B[e] Phenomenon. ASP
Conf. Series, Vol. 355, p.197
\bibitem{} Price S.D., Egan M.P., Carey S.J., Mizuno D.R., Kuchar T.A., 2001, AJ, 121, 2819
\bibitem{} Puls J. et al., 1996, A\&A, 305, 171
\bibitem{} Puls J., Vink J.S., Najarro F., 2008, A\&A Rev., 16, 209
\bibitem{} Puls J., Urbaneja M.A., Venero R., Repolust T., Springmann U., Jokuthy A., Mokiem M.R., 2005, A\&A, 435, 669
\bibitem{} Reid M.J., Menten K.M., Zheng X.W., Brunthaler A., Xu Y., 2009, ApJ, 705, 1548
\bibitem{} Reipurth B., Schneider N., 2008, in ASP Monograph Publications, Vol. 4, Handbook of Star Forming Regions, Vol. I:
The Northern Sky, ed. B.Reipurth (San Francisco: ASP), 36
\bibitem{} Rieke G.H. et al., 2004, ApJS, 154, 25
\bibitem{} Rivero Gon\'{z}alez J.G., Puls J., Najarro F., 2011, A\&A, 536, A58
\bibitem{} Ryans R.S.I., Dufton P.L., Rolleston W.R.J., Lennon D.J., Keenan F.P., Smoker J.V., Lambert D.L., 2002, MNRAS, 336, 577
\bibitem{} Rygl K.L.J. et al., 2012, A\&A, 539, A79
\bibitem{} Santolaya-Rey A.E., Puls J., Herrero A., 1997, A\&A, 323, 488
\bibitem{} Schaerer D., Schmutz W., 1994, A\&A, 288, 231
\bibitem{} Schilbach E., R\"{o}ser S., 2008, A\&A, 489, 105
\bibitem{} Schneider N., Bontemps S., Simon R., Jakob H., Motte F., Miller M., Kramer C., Stutzki J., 2006, A\&A, 458, 855
\bibitem{} Sch\"onrich R., Binney J., Dehnen W., 2010, MNRAS, 403, 1829
\bibitem{} Shevchenko V.S., Ibragimov M.A., Chernysheva T.L., 1991, SvA, 35, 229
\bibitem{} Smith N., 2007, AJ, 133, 1034
\bibitem{} Sota A., Ma\'{i}z-Apell\'{a}niz J., Walborn N.R., Alfaro E.J., Barb\'{a} R.H., Morrell N.I., Gamen R.C., Arias J.I.,
2011, ApJS, 139, 24
\bibitem{} Steele I.A., Negueruela I., Clark J.S., 1999, A\&AS, 137, 147
\bibitem{} Stringfellow G.S., Gvaramadze V.V., Beletsky Y., Kniazev A.Y., 2012a, in Richards M.T., Hubeny I., eds, Proc. IAU Symp.,
Vol. 282, p. 267
\bibitem{} Takita S. et al., 2009, PASJ, 61, 291
\bibitem{} Tull R.G., MacQueen P.J., Sneden C., Lambert D.L., 1995, PASP, 107, 251
\bibitem{} Turner D.G., Forbes D., 1982, PASP, 94, 789
\bibitem{} Turner D.G., Rohanizadegan M., Berdnikov L.N., Pastukhova E.N., 2006, PASP, 118, 1533
\bibitem{} Urbaneja M.A., Herrero A., Kudritzki R., Najarro F., Smartt S.J., Puls J., Lennon D.J., Corral L.J., 2005, ApJ, 635, 311
\bibitem{} van Buren D., Noriega-Crespo A., Dgani R., 1995, AJ, 110, 2914
\bibitem{} van Kerkwijk M.H., van Oijen J.G.J., van den Heuvel E.P.J., 1989, A\&A, 209, 173
\bibitem{} Vansevi\v{c}ius V., Brid\v{z}ius A., Pu\v{c}inskas A., Sasaki T., 1996, Baltic Astron., 5, 539
\bibitem{} Wachter S., Mauerhan J., van Dyk S., Hoard D. W., Morris P., 2011, Bull. Soc. R. Sci. Li\`{e}ge, 80, 291
\bibitem{} Wachter S., Mauerhan J.C., van Dyk S.D., Hoard D.W., Kafka S., Morris P.W., 2010, AJ, 139, 2330
\bibitem{} Walborn N.R., Fitzpatrick E.L., 1990, PASP, 102, 379
\bibitem{} Wang J.-J., Hu J.-Y., 2000, A\&A, 356, 118
\bibitem{} Weaver R., McCray R., Castor J., Shapiro P., Moore R., 1977, ApJ, 218, 377
\bibitem{} Werner M.W. et al., 2004, ApJS, 154, 1
\bibitem{} Wright E.L. et al., 2010, AJ, 140, 1868
\bibitem{} Zacharias N., Finch C.T., Girard T.M., Henden A., Bartlett J.L., Monet D.G., Zacharias M.I., 2013, AJ, 145, 44

\end{thebibliography}
\end{document}